# Multi-delay arterial spin-labeled perfusion estimation with biophysics simulation and deep learning

Renjiu Hu[1,3], Qihao Zhang[3], Pascal Spincemaille[3], Thanh D. Nguyen[3], Yi Wang[2,3]*

[1]*Sibley School of Mechanical and Aerospace Engineering, Cornell University, Ithaca, New York, USA*
[2]*Meinig School of Biomedical Engineering, Cornell University, Ithaca, New York, USA*
[3]*Department of Radiology, Weill Medical College of Cornell University, New York, New York, USA*

* Corresponding author: yiwang@med.cornell.edu

**ABBREVIATIONS:**

$Q$ = perfusion (mL/100g/min)

$NRMSE$ = normalized root mean squared error

# ABSTRACT

**Purpose:** To develop biophysics-based method for estimating perfusion $Q$ from arterial spin labeling (ASL) images using deep learning.

**Methods:** A 3D U-Net (QTMnet) was trained to estimate perfusion from 4D tracer propagation images. The network was trained and tested on simulated 4D tracer concentration data based on artificial vasculature structure generated by constrained constructive optimization (CCO) method. The trained network was further tested in a synthetic brain ASL image based on vasculature network extracted from magnetic resonance (MR) angiography. The estimations from both trained network and a conventional kinetic model were compared in ASL images acquired from eight healthy volunteers.

**Results:** QTMnet accurately reconstructed perfusion $Q$ from concentration data. Relative error of the synthetic brain ASL image was 7.04% for perfusion $Q$, lower than the error using single-delay ASL model: 25.15% for $Q$, and multi-delay ASL model: 12.62% for perfusion $Q$.

**Conclusion:** QTMnet provides accurate estimation on perfusion parameters and is a promising approach as a clinical ASL MRI image processing pipeline.

**INTRODUCTION**

Quantitative transport mapping (QTM) method is a recently developed physics-based perfusion quantification method using fluid dynamics to model tracer transport[1-3]. QTM estimates voxel averaged transport velocity |u| by solving the inverse problem of the continuity equation with total variation regularization[4]. QTM not only overperformed the traditional kinetic models in simulated arterial spin labeling (ASL) images[4, 5], but also showed a higher diagnostic performance in distinguishing benign and malignant breast lesions from dynamic contrast-enhanced images than the vascular exchange rate $K^{trans}$ and extravascular extracellular space $V_e$ calculated from conventional kinetic model[6, 7].

However, to estimate the perfusion $Q$ via QTM method, either the knowledge of the local vascular system should be obtained[4] or the homogeneous porous media feature should be hold[8]. For organs with more complex vasculatures such as brain, the previous prerequisites can hardly be hold and an alternative approach needs to be established for the universal perfusion prediction.

Deep learning is a rapid growth and widely used method for the data analysis in both physics and medical imaging fields. In the physics filed, deep learning has been used for fluid simulation acceleration[9], fluid velocity field estimation[10], and interface structure selection for optimal heat flow[11]; In the medical imaging field, deep learning has already been a popular method in rim classification[12, 13], multiple sclerosis lesion segmentation[14], brain registration[15, 16], and perfusion image generation[17-19]. Moreover, deep learning has remarkable performance in transfer learning from synthetic data to clinical data in registration and segmentation[16, 20]. The excellent performance of deep learning in physics simulation, medical imaging processing and transfer learning makes it a reliable candidate in estimating the perfusion of the ASL images by learning the fluid patterns from the simulated data.

In this work, we proposed QTMNet, a biophysics-based deep learning QTM solver, to estimate the perfusion from ASL images. We used constrained constructive optimization (CCO) [21] to numerically construct 4D tracer propagation data for training and testing. The network performance was evaluated on simulated brain perfusion model based on a MR angiogram images as well as in ASL images acquired from eight healthy volunteers.

## METHODS

**Theory**

In vascular space, the convection diffusion equation could be used to describe the tracer propagation:

$$\frac{\partial c_p(r,t)}{\partial t} = \nabla \cdot \left(-\boldsymbol{u}(r)c_p(r,t) + D(r)\nabla c_p(r,t)\right) - \gamma c_p(r,t), \quad (1)$$

Here $c_p(r,t)$ is the tracer concentration at coordinate $r$ and time $t$ in the vascular space, $\boldsymbol{u}(r)$ is velocity and $D(r)$ is diffusion coefficient which is 0 due to the negligible effect[4]. $\gamma = 1/T_1$ with $T_1 = 1650\ ms$ for labeled tracer in ASL on 3T scanner[22].

The vascular space was divided into three mutually exclusive subspace arterial (arterial vessels with radii > 6$\mu$m), capillary (vessels with radii ⩽6$\mu$m) and venous (venous vessels with radii > 6$\mu$m) for modeling. By taking the volume integral of Eq. 1 in the vascular and using the divergence theorem, we obtain:

$$\frac{\partial \iiint_{V_a(\xi)} c_p(r,t)dV}{\partial t} + \gamma \iiint_{V_a(\xi)} c_p(r,t)dV$$

$$= \oiint_{V_a \cap VS(\xi)} \left(-\boldsymbol{u}(r)c_p(r,t) + D(r)\nabla c_p(r,t)\right)d\vec{S} + \oiint_{V_a \cap V_c(\xi)} \left(-\boldsymbol{u}(r)c_p(r,t) + D(r)\nabla c_p(r,t)\right)d\vec{S}, \quad (3)$$

and

$$\frac{\partial \iiint_{V_v(\xi)} c_v(r,t)dV}{\partial t} + \gamma \iiint_{V_v(\xi)} c_p(r,t)dV$$

$$= \oiint_{V_v \cap VS(\xi)} \left(-\boldsymbol{u}(r)c_v(r,t) + D(r)\nabla c_v(r,t)\right)d\vec{S} + \oiint_{V_v \cap V_c(\xi)} \left(-\boldsymbol{u}(r)c_v(r,t) + D(r)\nabla c_v(r,t)\right)d\vec{S}, \quad (4)$$

Here $V_a$, $V_c$, and $V_v$ denote the volume of arterial, capillary, and venous subspace, respectively, and $VS$ is voxel surface. In the integral, the positive direction of surface $\vec{s}$ is towards the outside of the corresponding volume. $V_i \cap V_j(\xi)$ denotes the contacting surface of two exclusive subspace at voxel $\xi$, $V_i \cap VS(\xi)$ denotes the cross-section of the vascular volume and voxel surfaces at voxel $\xi$, and $\cup V_i(\xi)$ denotes the union of volumes at voxel $\xi$ (For example, $V_a \cap VS(\xi)$ represents the cross section of the arterial subspace and the voxel surface, $V_a \cup V_c \cup V_v(\xi)$ represents total blood volume at voxel $\xi$). We define the perfusion $Q$ of the voxel $\xi$ as the flow from the arterial space to the capillary space and the amount of tracer C in the voxel $\xi$ is the integration of concentration in the corresponding voxel:

$$C(\xi, t) = \oiiint_{V_a \cup V_c \cup V_v(\xi)} c_v(r, t) dV, \qquad (5)$$

$$Q(\xi) = \oiint_{V_a \cap V_c(\xi)} \boldsymbol{u}(r) d\vec{\boldsymbol{S}}, \qquad (6)$$

The corresponding inverse problem is to solve $Q(\xi)$ from $C(\xi, t)$ based on the equations above, which is highly nonlinear and vascular dependent. Therefore, we propose the new method using a neural network (QTMnet) to estimate $Q(\xi)$ from $C(\xi, t)$ via learning the perfusion characteristics from the artificial biophysical simulated data.

**Data simulation**

We utilize the anatomic information of the brain vessels to generate the brain vessel system. For the vessel system construction, all the vessels were assumed to be cylindrical with laminar flow running through, thus the flow velocity profile is parabolic. The large vessels were firstly extracted from MR angiogram data and the CCO method was later applied on the large vessel to generate the pial vessels. These two formed the level 1 vessel in cerebrospinal fluid (CSF) region. Then, the arterioles/venules were built in the grey matter (GM) and white matter (WM) regions to create level 2 vessel system. Finally, the capillary part was built to connect the arterioles and venules within the same voxels. Algorithm details can be found in Appendix. The radii and flow information were stored in the edge matrix and the position information was stored in the vortex matrix for further simulation.

We applied the fluid dynamics along with the perfusion parameters to build a brain perfusion model. In the fluid dynamics simulation, the initial values of the $Q$ was preset using the literature values and the detailed values can be found in Appendix. To increase the generality, the initial values would have 10% fluctuation around the average values. With the knowledge of the CBF at the terminal, the blood flow inside the vessels system could be calculated through the superposition of the blood flow from the daughter vessels. The labeling flow started from internal carotid arteries and basilar artery and the arterial input function could be determined by the following equation:

$$AIF(x,t) = [H(x, t - ATT(x)) - H(x, t - ATT(x) - LD)] \exp\left(\frac{t_{arr}(x)}{T_1}\right), \tag{1}$$

Here $H(x,t)$ is the Heaviside function at position $x$, $ATT(x)$ is the arterial transit time of the tracer at position $x$, LD is the labeling duration (1450 ms in the simulation) and $t_{arr}(x)$ is the arriving time of the tracer at position $x$ to simulate the tracer decaying under 3T. The arterial transit time of every vortex was calculated recursive and stored in an additional matrix:

$$ATT(x) = \frac{l}{\frac{Q}{\pi r^2}} + ATT(x_{-1}), \tag{2}$$

Here, $ATT(x)$ denotes the arterial transit time of the vortex $x$, $x_{-1}$ denotes the father node of the vortex $x$, and $\frac{l}{Q/\pi r^2}$ denotes the additional transit time from $x_{-1}$ to $x$. $l$ is the segment length between the two vortices which could be calculated through the stored position information. $Q$ and $r$ are the flow and radius respectively.

The brain perfusion model was voxelized to generate the simulated ASL images. The Cartesian coordinates $x$ was rounded up and divided by the spatial resolution ($1 \times 1 \times 1\ mm^3$ in this case) to get the corresponding voxel coordinates $\xi$. Later, the finite-element version two-compartment exchange model was applied to every voxel to simulate the tracer exchange in the capillary system. The simulation temporal resolution is 10ms and the image capture temporal step is 0.5s.

We used artificial vessel cubes for the network training. We created $32 \times 32 \times 32$ cubes for vessel generation. The region growth algorithm was used to randomly generate different regions within the cubes and different perfusion parameters were preset in different regions using literature values. The local two-compartment exchange model was then used to simulate the tracer exchange in the capillary area. More details can be found in Appendix.

All the simulations were finished on a server with Intel i9-9940X 14-core CPU and 128 GB memory.

**Data acquisition**

We applied Pseudo-continuous ASL (PCASL) 3D fast spin echo (FSE) sequence to acquire data of the brain in healthy volunteers (N = 8; five males, three females) using a GE MR750 3T scanner (GEHC, Milwaukee, WI) with a 32-channel head coil. The acquisition parameters were: image voxel size $1.875 \times 1.875 \times 4\ mm^3$, image volume size $128 \times 128 \times 36$, 10.5 ms echo time, three signal averages ($NEX = 3$), 1450 ms labeling duration. Four post-labeling delays (PLD = 1025 ms, 1525 ms, 2025 ms, 2525 ms) were used during the data acquisition to obtain different ASL images. Background suppression was used to enhance the signal quality.

**Network architecture**

QTMNet was based on a 3D U-Net [23] with the modification on the bottleneck as exhibited in Figure 1. The channel-spatial attention module[24] has been appended to the two convolution operations in the bottleneck to increase the reconstruction performance. The temporal dimension replaced the normal channel dimension to make all the time frames as the input. Batch normalization[25] was removed from the architecture as it would deteriorate the reconstruction performance.

**Network training**

QTMNet was trained on the artificial vessel cubes to learn the patterns of the flow with the time changes. The total 6400 vessel cubes were split into training and validation set on 9: 1 ratio and they were normalized by the global maximum intensity of the cubes. The training epochs were 40. We used stochastic gradient optimizer with learning rate 1e-3 and momentum 0.9, and the cosine learning rate scheduler was applied to reduce the learning rate smoothly with the increase of the training echo until 1e-6. The L1 loss along with the gradient penalty on blood flow was used as the loss function.

$$\mathcal{L}(\Psi) = \|Q - \Psi(C)\|_1 + \lambda\|\nabla\Psi(C)\|_1$$

Here $\Psi$ represents the QTMNet. The network training was finished on Nvidia GeForce RTX 2080.

**Image analysis**

For the brain perfusion model, we downsampled the image voxel to $2 \times 2 \times 4$ to match the typical real ASL image voxel size. To be evaluated by QTMNet, the simulated images were normalized by the global maximum intensity of the brain perfusion model. On the same time, the traditional Kety's models were also applied for comparison. For single delay ASL, only the ASL image with $PLD = 1525\ ms$ was used for the evaluation and the following equation was used[22]:

$$CBF_{single} = \lambda \left(1 - \exp\left(-\frac{ST}{T1_t}\right)\right) \frac{\exp\left(\frac{PLD}{T1_b}\right)}{2T1_b\left(1 - \exp\left(-\frac{LT}{T1_b}\right)\right)\epsilon} \frac{dM}{NEX_{PW} SF_{PW} M_0} \quad (6)$$

Here $CBF_{single}$ represents the predicted CBF from the single delay ASL, $ST = 2\ s$ is the saturation time, PLD is the post-labeling delay, $T1_t = 1.2\ s$ and $T1_b = 1.65\ s$ is the $T_1$ of tissue and blood correspondingly. $LT = 1450\ ms$ is the labeling time, dM is the ASL image and $M_0$ is the proton density-weighted image. $NEX_{PW} = 3$ is the number of excitations for ASL images and $SF_{PW} = 32$ is the scaling factor of the ASL sequence and both are specifically used for GE scanner. The partial saturation correction factor $\left(1 - \exp\left(-\frac{ST}{T1_t}\right)\right)$ along with the two GE-specific factors were only used when estimating the health volunteer data and they would be set as 1 when dealing with the simulated data. $\lambda = 0.9$ is the blood/tissue partition coefficient and $\epsilon = 0.6$ is the labeling efficiency, and both were set as 1 when dealing with the simulated data. Also, the factor 2 in the denominator would also be set as 1 because there was no spin flip in the simulation.

For multi delay ASL, the following equation was used for the evaluation[4, 26, 27]:

$$\lambda dM(PLD) = CBF_{multi} * 2\alpha M_0 T_1' \exp\left(-\frac{ATT}{T1_b}\right)\left[1 - \exp\left(-\frac{\min(PLD - ATT, LT)}{T_1'}\right)\right] \exp\left(-\frac{\max(0, PLD - ATT - LT)}{T_1'}\right) \quad (7)$$

Where $CBF_{multi}$ represents the predicted CBF from the multi delay ASL, $\frac{1}{T_1'} = \frac{1}{T1_t} + \frac{CBF_{multi}}{\lambda}$ is the modified $T_1$ of tissue and $ATT$ is the arterial transit time of blood from labeling location to the brain estimated from the weighted delay[4, 26, 27]. All the frames (i.e., $PLD = 1025, 1525, 2025, 2525\ ms$) would be used for the evaluation.

RESULTS

The retrospective analysis of ASL data in this study was approved by the Institutional Review Board and was HIPAA compliant.

**Perfusion parameter estimation accuracy of QTMnet and kinetic modeling**

Figure 2d-2g are the simulated ASL images with different PLDs that were used for the evaluation. QTMnet predicted perfusion $Q$ from a concentration profile on the brain perfusion model is shown in Figure 3d. As a comparison, ground truth, single-delay ASL and multi-delay ASL model predicted perfusion $Q$ are simultaneously shown in Figure 3a, 3b and 3c, accordingly. The NRMSE of QTMnet predicted perfusion parameter is 7.04% for $Q$, while the NRMSE of single- or multi-delay ASL predicted perfusion parameter is 25.15% and 12.62% for $Q$, respectively. From the corresponding error maps, we could see that the single-delay ASL method has large errors in posterior cerebral area; multi-delay ASL method mitigates this problem by leveraging the signals under different PLDs, but the voxel-based method cannot fully eliminate the errors introduced by various transit time comparing to the QTMnet method, as exhibited in Figure 3e-3f. Transfer learning from the artificial synthetic training data assists the QTMnet to leverage the patterns of longer-transit-time blood of the posterior cerebral region to generate a more uniform perfusion flow map.

The prediction on the healthy volunteer has been shown in Figure 4. QTMnet similarly provided a more uniform perfusion $Q$ of the gray matter in posterior cerebral region compared to the traditional kinetic method. Figure 5 exhibits that on both simulation and healthy volunteer data, the QTMNet improves the prediction of flow on the posterior region.

## DISCUSSION

We present a biophysics-base method that using simulated tracer propagation in artificial micro-vasculature as training data and a neural network (QTMnet) as the inverse-problem solver. This method showed a 45% smaller estimation error compared to the traditional kinetic model on perfusion $Q$ from 4D tracer propagation at different resolutions with noise.

The consideration of the tracer transport in both macro- and micro-vasculature facilitate the new method become a more accurate model compared to the kinetic modeling method, where the transit delay and dispersion of tracer are neglected. To achieve this, the parabolic flow assumption, which is common in pipe flow, was employed inside vasculature and convection equation was also adopted to solve the tracer propagation over time. Similar approach has been applied to investigate the contrast agent and oxygen transport in normal tissue and tumor already [28].

Our proposed inverse-problem solver (QTMnet) overcomes the optimization difficulty related to the nonlinearity feature of the problem. The complexity of the vasculature and extravascular space in voxel forces the explicit expression of perfusion parameter requires the knowledges of vessel structure, which is unknown in most situations[4]. Nonetheless, the ignorance of the vasculature make the current bio-physics simulation degenerated to traditional kinetic model, which amplifies the delay and dispersion error [29]. QTMNet could appropriately overcome these difficulties by directly learning the connections between perfusion parameters and tracer concentration with diverse vasculature and perfusion parameter distributions. In the evaluation, the network can directly predict perfusion parameters from tracer concentration without prior vasculature information. QTMnet showed good perfusion parameter reconstruction accuracy in the simulated perfusion model than traditional kinetic methods.

The close-to-reality tracer propagation simulation and vasculature generation along with good neural network generalization enable the high performance of QTMnet in ASL processing. The CCO-method-based vasculature generation could generate vasculature statistically close to real vasculature given certain value[30]. Moreover, QTMnet is robust to noise and resolution change. These results indicate that QTMnet is promising to process various kinds of ASL images.

This bio-physics-based method could easily be extended to manage other perfusion-related problems. For instance, it could be used to reconstruct perfusion parameters from proton emission tomography (PET) or Dynamic contrast enhanced MRI acquisitions by modifying some tracer properties like specific uptake rate or longitudinal relaxation rate. After, it could also be applied to illustrate tracer propagation in different organs by adjusting the perfusion parameter value ranges and the constrains of the CCO method. Additionally, it could also be used to investigate other vessel related properties, such as vessel size distribution, oxygen extraction fraction, etc.

There exist several limitations in this study. Firstly, in the fluid dynamics simulation, the fluid was assumed to be well-developed, which guarantees the parabolic velocity profile. However, there would be some deviation at the vessels sub-branches[8], and the accumulation of the errors from the large vessels to the capillaries could affect the arterial transit time alongside the parabolic shape of the fluid velocity profile. Secondly, the simulation ignored the exchange of tracer between vascular and extravascular space, assuming a single compartment inside each voxel, which was less accurate in approximating the real brains. Finally, QTMnet has not been validated on perfusion phantom with known ground truth. This validation can further examine the accuracy of QTMnet in perfusion images.

In conclusion, we applied the biophysics-based simulation on artificial vessel structure and proposed a neural network (QTMnet) to reconstruct perfusion flows from 4D tracer concentration via transfer learning. QTMnet can reconstruct perfusion flows from 4D tracer concentration image such as ASL images with higher accuracy than traditional kinetic models. In future studies, QTMnet will be applied to processing of different perfusion weighted imaging modalities, such as PET.

**FUNDING SUPPORT**

**CONFLICT OF INTEREST DISCLOSURES**

**FIGURES**

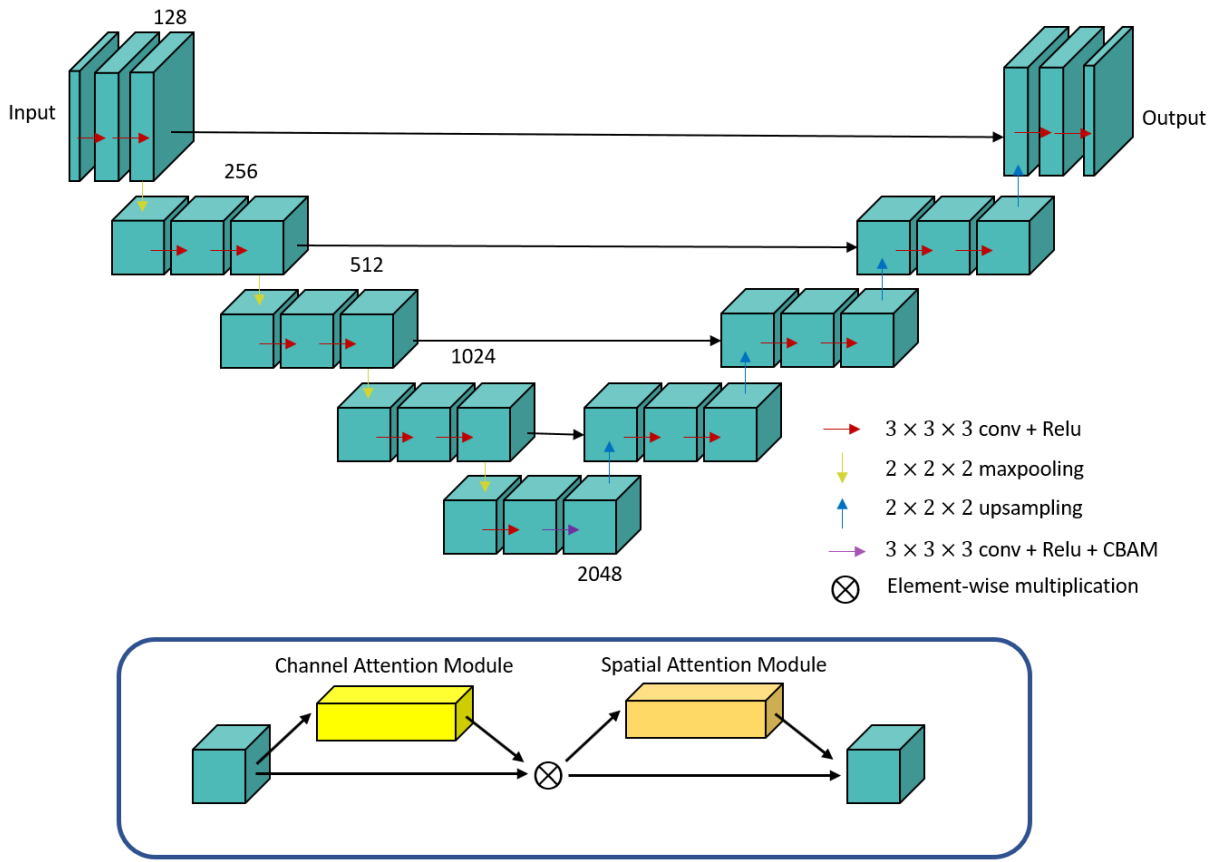

Figure 1. 3D U-net architecture for perfusion parameter reconstruction with CBAM appended in the bottleneck.

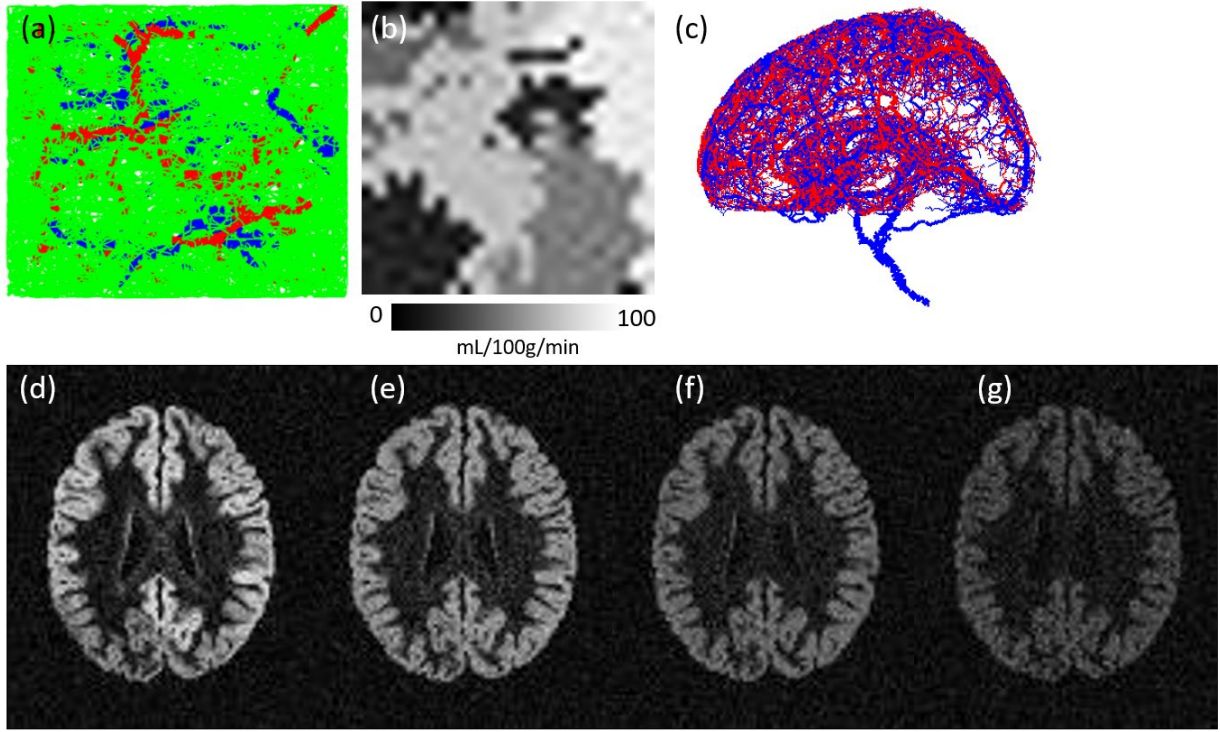

Figure 2. Data simulation of the artificial vessel cubes and brain perfusion model. a) artificial vessel cube generated via CCO method; b) the randomly preset perfusion. c) brain vasculature constructed by extracted vessels from MR angiography and CCO method. d) to g) are the simulated ASL concentration images with $LD = 1450 ms$ and $PLD = 1025, 1525, 2025, 2525 ms,$ respectively.

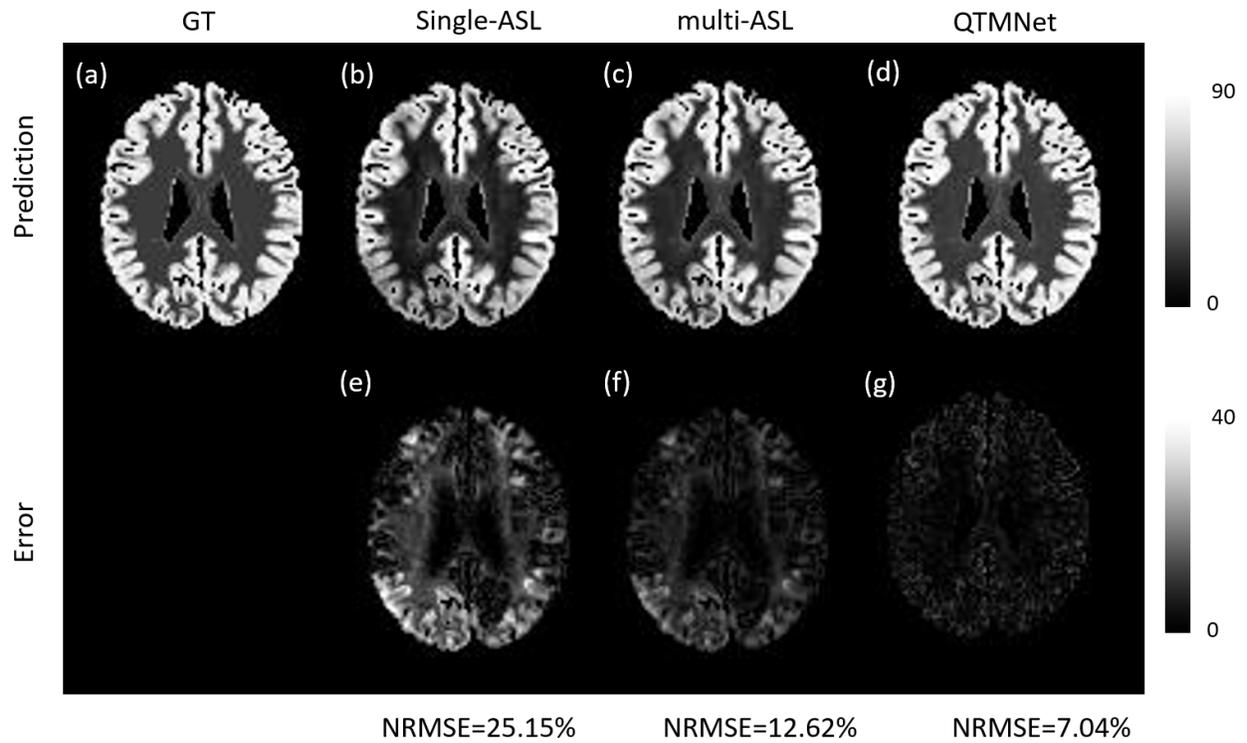

Figure 3. a) The ground truth perfusion $Q$ of the simulated brain perfusion model; b) – d) are the predicted perfusion via single-delay ASL model, multi-delay ASL model and QTMNet; e) – f) are the error maps between predicted perfusion and the ground truth

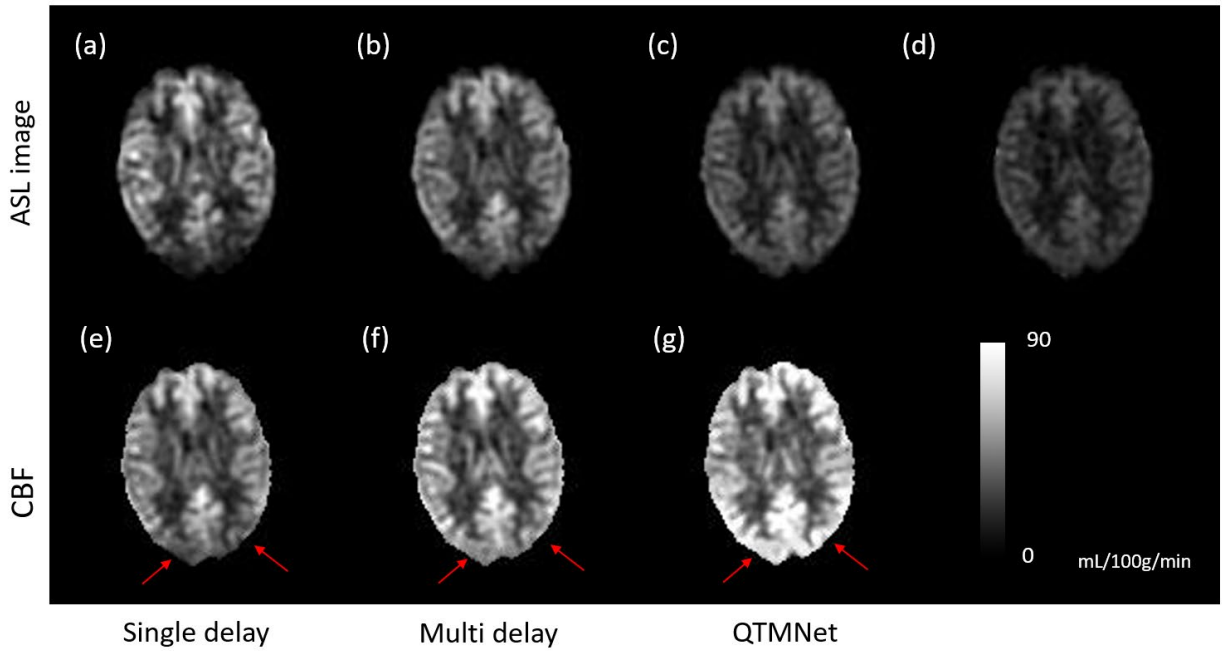

Figure 4. a) – d) are the ASL images from one healthy volunteer with $PLD = 1025, 1525, 2025, 2525 ms$. e) – g) The predicted $Q$ through single-delay ASL method, multi-delay ASL method and QTMnet, respectively.

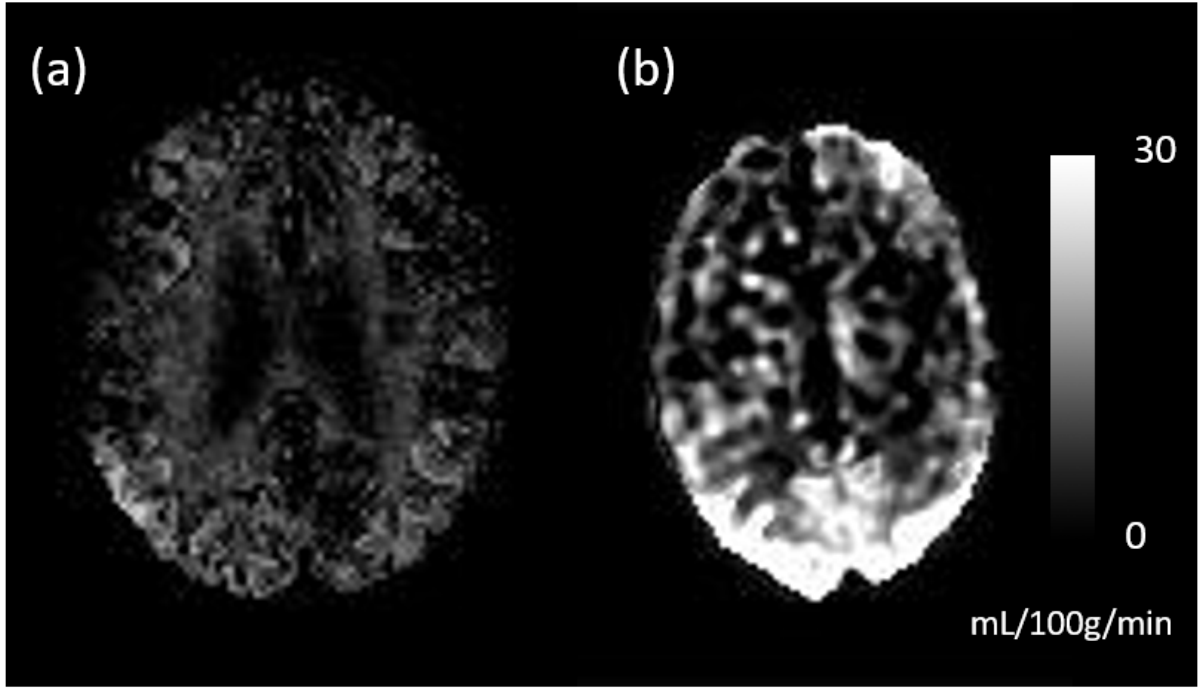

Figure 5. a) and b) are the deviation map between QTMNet and multi-delay ASL model predictions on perfusion $Q$ of the simulated and the healthy volunteer data, respectively.

# APPENDIX 1

**Artificial vessel cube generation based on constrained constructive optimization (CCO) method**

We considered the vasculature and tracer propagation inside a $32 \times 32 \times 32$ mm$^3$ volume. We used region growth algorithm to generate different regions. Perfusion $Q$ was assumed to be region-based.

In the simulation, $N = 40$ seeds were randomly set in the volume for the region growth. Every time a random seed would be selected for the region growing until it met the boundaries of the volume or other regions. The perfusion values from zero to one were randomly set for every region and for every region the values were uniform. Afterwards, Q were scaled to 0-80 mL/100g/min based on the literature values[31] with 10% fluctuation to increase the generality. Representative generated Q maps are shown in Figures 2b. After the parameter maps were settled, the vascular system (arteries, capillaries and veins) was built based on constrained constructive optimization (CCO) method[21]. The total volume was divided into $32 \times 32 \times 32$ grids with size $1 \times 1 \times 1$ mm$^3$. The construction pipeline is as follows. First, we define the following constants:

| | |
|---|---|
| $N_1 = 0.01$ | Probability that new node will start a new vascular tree instead of connecting to the existing tree |
| Terminal point | Outlet of the vascular tree, and will be connected to capillaries in the next step |
| $d_1 = 5$mm | The maximum length of vessel segment |
| $N_a = 3$ | Number of candidate segments that a terminal point will be connected to |
| $\epsilon(P)$ | Grid that contains vessel point P |
| $Q(\epsilon)$ | Flow of grid $\epsilon$, acquired from the previous F generation step |

and the following three criteria:

| | |
|---|---|
| $C_1(P)$ | No other terminal exists inside the same grid P belongs to. |
| $C_2(S, P)$ | The distance between vessel segment S and P is smaller than $d_1$ |

$C_3(P_1, P_2)$ The distance between $P_1$ and $P_2$ is smaller than $d_1$, and $P_1$ and $P_2$ are not in the same grid

Then the algorithm for the arterial tree can be written as

```
1   Do:
2       Randomly draw a candidate terminal point P_1 inside the volume such that C_1(P_1) is true.
3       If rand(1) > N_1
4           Find N_a vessel segments S_i closest to P_1 such that C_2(S_i, P_1) is true.
5           For each vessel segment S_i found in line 4
6               Set flow of terminal point F_1 as F(ϵ(P_1)) and connect to S_i.
7               Update the flow of vascular tree.
8               Determine the optimal connecting point P_2 from P_1 to S_i by minimizing the total
                vessel tree volume.
9           Traverse possible vessel segment S_i and connect P_1 to the vessel segment that gives the
            minimal total vessel tree volume.
10          Update the flow of vessel tree.
11      Else
12          Set P_1 as another arterial inlet.
13          If node P_2 for which C_3(P_1, P_2) is true can be drawn in the volume:
14              Connect P_2 to P_1 as the first segment of this new artery
                Set flow of P_2 to F_2 from perfusion map Q.
15          Else
16              Discard P_1.
17          End
18      End
19  Until there is no more terminal point P such that C_1(P) is true.
```

While minimizing the total vasculature volume, we assumed the cubic of vessel diameter is proportional to flow [32, 33]. The proportionality factor was set to one, as the relative vessel volume was helpful enough for the vessel structure optimization. The arterial and venous vasculatures were constructed separately using the same method, and representative generated arterial and venous vasculatures are shown in Figure 2a.

The above algorithm guarantees that there exists only one terminal artery and terminal vein within each grid $\epsilon$ with the same flow $Q(\epsilon)$. Capillaries were then constructed to connect the terminal artery-vein pairs. We further assumed that the capillaries stayed within the same grid as the terminal artery/vein, and they could only connect the pair inside the same grid. Hence, only the length, radii, and velocity of the capillaries need to be considered instead of the morphology and bifurcation information. In the construction, a random integer $N_{cap}$ between 50 and 150 from a uniform distribution was chosen as the number of capillaries inside the grid. Every capillary has the radius, length, and flow be set with random values between 0.5 and 1.5 from a uniform distribution. These parameters were later being scaled to match the constrains of the total flow $Q(\varepsilon)$. In this way, a fully connected vascular network was constructed. With the knowledge of the terminal flow along with the flow direction, the flow and velocity inside each vessel segment could be determined.

**Vascular generation for brain**

Arteries and veins were extracted from MR angiography images of a brain registered to MNI standard space and were digitalized into unidirectional trees consisting of vortices and edges with certain radii and length, which is shown in Figure 2c. Limited by image resolution, most of the vessel radii are larger than 1 mm, lacking the feeding arterioles, draining veins and capillaries. Therefore, these small vessels were constructed with the CCO method and perfusion map $Q$ were assumed to be uniform in different region (i.e., WM and GM) with literature values as described in previous section.

# APPENDIX 2

**Tracer propagation simulation based on vasculature and perfusion parameters**

The following three assumptions was made during the tracer propagation simulation in the vascular system:

(1) The vessel wall permeability of arteries and veins are neglectable.

(2) The diffusion effect inside arteries and veins are neglectable.

(3) Each vessel is a cylinder. The flow in capillaries is plug flow and the flow in arteries and veins is well-developed parabolic flow.

The mass exchange of blood and tissue mostly happen in capillaries[34]. Assumption 1 allows us to simulate the flow in artery, capillary, and veins separately by making the output of the previous part the input of the next part. The diffusion coefficient of water is $1 \times 10^{-3}$ mm$^2$/s, and the average displacement by diffusion is several magnitude smaller than the convectional transport of blood[35]; Plug flow assumption in capillaries and parabolic flow assumption in arteries and veins are validated in a previous study[36]. Assumption 2 and 3 allows a closed form solution for tracer propagation in arteries and veins. For the arterial component, within one segment, the concentration c at each point could be stated as:

$$c(x, r, t) = c\left(0, r, t - \frac{x}{u(r)}\right), \quad (A1)$$

Here x is the distance away from the starting point of the vessel segment along the positive direction of the flow, r is the radical distance away from the original point of the cylinder and u is the velocity. At the bifurcation point, the concentration is preserved:

$$c_d(0, r, t) = c_f\left(l_f, \frac{R_f}{R_d} r, t\right), \quad (A2)$$

where subscripts f and d refer to father and daughter branches, respectively. In arterial side, father branch means the branch in upstream direction, while in venous side father branch means the opposite. $l_f$ is the length of the father segment. The formation of the concentration input was mentioned in the main content, see Eq (1). The concentration at the outlet is used as the inlet of capillaries for the capillary simulation.

For capillary simulation, finite element method was used for the tracer propagation simulation as there is no closed form solution. For capillaries, the vessel segments consist of 10 elements along the propagation direction. The concentration propagation between capillaries and EES is determined by following equations:

$$\frac{\partial V(\xi_c) c(\xi_c, t)}{\partial t} = u(\xi_c - 1) S_{\xi_c \cap \xi_c - 1}(\xi_c - 1) c(\xi_c - 1, t) - u(\xi) S_{\xi_c \cap \xi_c + 1}(\xi_c) c(\xi_c, t), \quad (A3)$$

Here $V(\xi_c)$, $u(\xi_c)$, $c(\xi_c, t)$ are the volume, velocity, and concentration of capillary element $\xi$, respectively. $\xi - 1$ and $\xi + 1$ are the neighboring element at the upstream/downstream direction of element $\xi$, correspondingly. $S_{\xi_c \cap \xi_c \pm 1}$ is the cross section of $\xi$ with neighboring voxels. The concentration at the outlet is further served as the inlet of veins for the venous tracer propagation simulation.

For the venous vasculature, the parabolic flow assumption is kept. While Eq. A1 still holds inside one vessel segment, at the branching point, we assume the concentration of father branch is the flow-weighted average of concentration in daughter branches:

$$c_f(0, r, t) = \sum_{d \in D(f)} \frac{Q_d}{Q_f} c_d \left( l_d, \frac{R_d}{R_f} r, t \right), \quad (A4)$$

Here, $Q_d$ and $Q_d$ are flows from the daughter and father vessel segment, and $l_d$ is the length of the father vessel segment. After the element-wise concentration simulation, we down-sample the concentration into voxel level by summing up concentration multiplied by volume for all elements in the voxel. Simulated concentration profiles were down-sampled to $2 \times 2 \times 4\ mm^3$ voxel size with 0.5s temporal resolution.